\documentclass[12pt,oneside,submit]{JCNtran}

\usepackage[dvips]{epsfig}
\usepackage[latin1]{inputenc}
\usepackage[T1]{fontenc}
\usepackage{enumerate}
\usepackage{tabularx}

\begin{document}
\bibliographystyle{jcn}

\title{Cloud Radio Access Network: Virtualizing Wireless Access for Dense Heterogeneous Systems}
\author{Osvaldo Simeone, Andreas Maeder, Mugen Peng, Onur Sahin, and Wei Yu
\thanks{The work of O. Simeone has been partly supported by U.S. NSF under grant CCF-1525629}
\thanks{O.~ Simeone is with New Jersey Institute of Technology, NJ, USA, email: osvaldo.simeone@njit.edu.}\thanks{A. Maeder is with NOKIA Networks, Munich, Germany, email: andreas.maeder@nokia.com }\thanks{M. Peng is
with Beijing University of Posts and Telecommunications, China, email: pmg@bupt.edu.cn.}\thanks{O. Sahin is
with InterDigital, UK, email: onur.sahin@interdigital.com.}\thanks{W. Yu is
with University of Toronto, Canada, email: weiyu@comm.utoronto.ca.}}  \maketitle

\begin{abstract}
Cloud Radio Access Network (C-RAN) refers to the virtualization of base station functionalities by means of cloud computing. This results in a novel cellular architecture in which low-cost wireless access points, known as radio units (RUs) or remote radio heads (RRHs), are centrally managed by a reconfigurable centralized "cloud", or central, unit (CU). C-RAN allows operators to reduce the capital and operating expenses needed to deploy and maintain dense heterogeneous networks. This critical advantage, along with spectral efficiency, statistical multiplexing and load balancing gains, make C-RAN well positioned to be one of the key technologies in the development of 5G systems. In this paper, a succinct overview is presented regarding the state of the art on the research on C-RAN with emphasis on fronthaul compression, baseband processing, medium access control, resource allocation, system-level considerations and standardization efforts.
\end{abstract}

\section{\uppercase{Introduction}}

\label{sec:introd} Cloud Radio Access Network (C-RAN) refers to the virtualization of base station functionalities by means of cloud computing. In a C-RAN, the baseband and higher-layers operations of the base stations are implemented on centralized, typically general-purpose, processors, rather than on the local hardware of the wireless access nodes. The access points hence retain only radio functionalities and need not implement the protocol stack of full-fledged base stations. This results in a novel cellular architecture in which low-cost wireless access nodes, known as radio units (RUs) or remote radio heads (RRHs), are centrally managed by a reconfigurable centralized "cloud", or central, unit (CU). At a high level, the C-RAN concept can be seen as an instance of network function virtualization (NFV) techniques and hence as the RAN counterpart of the separation of control and data planes proposed for the core network in software-defined networking (see, e.g., \cite{Han}).

Referring to \cite{Checko} for a discussion of the origin and evolution of the C-RAN concept, we observe here that this novel architecture has the following key advantages:

\begin{itemize}

    \item It reduces the cost for the deployment of dense heterogeneous
networks, owing to the possibility to substitute full-fledged base
stations with RUs having reduced space and energy requirements;
    \item It enables the flexible allocation of radio and computing resources across all the connected RUs managed by the same CU, hence reaping statistical multiplexing gains due to load balancing;
    \item It facilitates the implementation of coordinated and cooperative transmission/ reception strategies, such as enhanced Inter-Cell Interference Coordination (eICIC) and Coordinated Multi-Point transmission (CoMP) in LTE-A, across the RUs connected to the same CU, thus boosting the spectral efficiency;
    \item It simplifies network upgrades and maintenance due to the centralization of RAN functionalities.

\end{itemize}

\begin{figure}[t]
\begin{center}
\epsfxsize=8cm \leavevmode\epsfbox{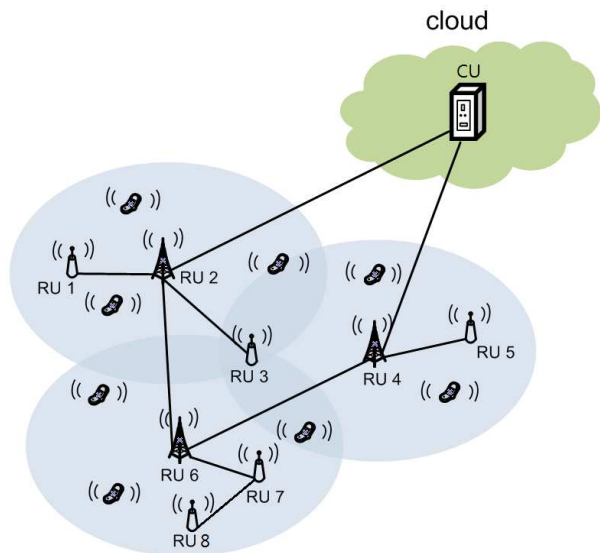} \caption{Illustration of a C-RAN with a general multi-hop fronthaul network.} \label{fig:uplink}
\end{center}
\end{figure}

In a C-RAN, as mentioned, the RUs implement only radio functionalities, including transmission/reception, filtering, amplification, down- and up-conversion and possibly analog-to-digital conversion (ADC) and digital-to-analog conversion (DAC). Therefore, for the downlink, each RU needs to receive from the CU either directly the analog radio signal - possibly at an intermediate frequency - that it is to transmit on the radio interface, or a digitized version of the corresponding baseband samples. In a similar fashion, in the uplink, the RUs are required to convey their respective received signals, either in analog format or in the form of digitized baseband samples, to the CU for processing. We refer to Fig. \ref{fig:uplink} for an illustration.
The RU-CU bidirectional links that carry such information are referred to as \textit{fronthaul} links, in contrast to the backhaul links connecting the CU to the core network. The analog transport solution is typically implemented on fronthaul links by means of radio-over-fiber (see, e.g., \cite{RoF}), but techniques based on copper LAN cables are also available \cite{RadioDots}. Instead, the digital transmission of baseband, or IQ, samples is currently carried out by following the CPRI standard \cite{CPRI}, which conventionally also requires fiber optic fronthaul links. The digital approach appears to be favored due to the traditional advantages of digital solutions, including resilience to noise and hardware impairments and flexibility in the transport options (see, e.g., \cite{Checko}).

The main roadblock to the realization of the mentioned promises of
C-RANs hinges on the effective integration of the wireless interface
provided by the RUs with the fronthaul transport network. In fact, the
inherent restriction on bandwidth and latency of the fronthaul links
may limit the effectiveness of cloud processing.
As an example, the latency induced by two-way fronthaul communication
may prevent the use of standard closed-loop error recovery techniques. These problems may be alleviated by a more flexible separation of functionalities between RUs and CU whereby parts of the baseband processing, such as FFT/IFFT, demapping and synchronization, and possibly of higher layers, such as error detection, are carried out at the RU \cite{Dotsch} \cite{Wubben}.

In this paper, we provide a brief overview of the state of the art on the research on C-RAN with emphasis on fronthaul compression, baseband processing, medium access control, resource allocation, system-level considerations and standardization efforts. We start in Sec. II with a discussion of typical baseband models used in the analysis of C-RAN systems. Then, in Sec. III, solutions for fronthaul transport and compression are reviewed. This is followed in Sec. IV by a review of relevant baseband processing techniques for C-RAN, along with the corresponding information theoretic analysis. Sec. V and Sec. VI cover design issues pertaining to higher layers, namely medium access layer and radio resource management, respectively. Sec. VII  elaborates on architectural considerations, and Sec. VIII provides a short discussion on standardization efforts. Finally, Sec. IX closes the paper with some concluding remarks.

\section{\uppercase{C-RAN Signal Models}}\label{sec:signal-model}

In order to introduce some of the main definitions, we start with a brief discussion of basic C-RAN signal models that are typically used in the analysis of the physical layer of C-RAN and that will be often referred to in the paper.

\subsection{Uplink}

In a C-RAN, the RUs are partitioned into clusters, such that all RUs
within a cluster are managed by a single CU. Within
the area covered by a given cluster, there are $N_{U}$ multi-antenna
user equipments (UEs) and $N_{R}$ multi-antenna RUs. In the uplink,
the UEs transmit wirelessly to the RUs. The fronthaul network connecting the RUs to the CU may have a single-hop
topology, in which all RUs are directly connected to the CU, or, more
generally, a multi-hop topology, as illustrated in Fig. \ref{fig:uplink}. An example of a single-hop
C-RAN is the network shown in Fig. 1 when restricted to RU 2 and RU
4.

Focusing for brevity on flat-fading channels, the discrete-time complex baseband,
or IQ, signal ${\mathbf{y}}_{i}^{\mathrm{ul}}$ received by the $i$th
RU at any given time sample can be written using the standard linear
model
\begin{equation}
{\mathbf{y}}_{i}^{\mathrm{ul}}={\mathbf{H}}_{i}^{\mathrm{ul}}{\mathbf{x^{\mathrm{ul}}}}+{\mathbf{z}}_{i}^{\mathrm{ul}},\label{eq:channel-uplink}
\end{equation}
where ${\mathbf{H}}_{i}^{\mathrm{ul}}$ represents the channel matrix
from all the UEs in the cluster toward the $i$th RU; $\mathbf{x}^{\mathrm{ul}}$
is the vector of IQ samples from the signals transmitted by all the UEs in the cluster;
and ${\mathbf{z}}_{i}^{\mathrm{ul}}$
models thermal noise and the interference arising from the other clusters. Note that in (\ref{eq:channel-uplink}),
and in the following, we do not denote explicitly the dependence of
the signals on the sample index in order to simplify the notation. We will provide further details on the system model in Sec. \ref{sec:baseband}.

We observe that the received signal is typically oversampled at the RUs (see Sec. \ref{sec:fronthaul compression}) and that the signal model (\ref{eq:channel-uplink}) can generally account also for oversampling in the simple case under discussion of flat-fading channels. Moreover, different assumptions can be made regarding the time variability of the channel matrices depending on mobility and transmission parameters.

In the single-hop topology, each RU $i$ is connected
to the CU via a fronthaul link of capacity $C_{i}$ bits/s/Hz. The
fronthaul capacity is normalized to the bandwidth of the uplink channel.
This implies that for any uplink coding block of $n$ symbols, $nC_{i}$
bits can be transmitted on the $i$th fronthaul link. In a multi-hop topology, an RU may communicate to the CU over a cascade of finite-capacity links.

\subsection{Downlink}

In the downlink, similar to the uplink, assuming flat-fading channels, each UE $k$
in the cluster under study receives a discrete-time baseband signal
given as
\begin{equation}
{\mathbf{y}}_{k}^{\mathrm{dl}}={\mathbf{H}}_{k}^{\mathrm{dl}}{\mathbf{x}}^{\mathrm{dl}}+{\mathbf{z}}_{k}^{\mathrm{dl}},\label{eq:channel-downlink}
\end{equation}
where $\mathbf{x}^{\mathrm{dl}}$ is the aggregate baseband signal
vector sample transmitted by all the RUs in the cluster; the additive noise
${\mathbf{z}}_{k}^{\mathrm{dl}}$
accounts for thermal noise and interference from the other clusters;
and the matrix ${\mathbf{H}}_{k}^{\mathrm{dl}}$ denotes the channel
response matrix from all the RUs in the cluster toward UE $k$. The fronthaul network can also be modelled in the same fashion as for the uplink. Further discussion can be found in Sec. \ref{sec:baseband}.

\section{\uppercase{Fronthaul Compression}}\label{sec:fronthaul compression}

In this section, we provide an overview of the state of the art on the problem of transporting digitized IQ baseband signals on the fronthaul links. We first review the basics of the CPRI standard in Sec. \ref{sec:CPRI}. Then, having identified the limitations of the scalar quantization approach specified by CPRI, Sec. \ref{sec:ptp compression} reviews techniques that have been proposed to reduce the bit rate of CPRI by means of compression as applied separately on each fronthaul link, i.e., via \textit{point-to-point} compression. Finally, in Sec. \ref{sec:network-aware compression}, advanced solutions inspired by network information theory are discussed that adapt the compression strategy to the network and channel conditions by means of signal processing across multiple fronthaul links.

\subsection{Scalar Quantization: CPRI}\label{sec:CPRI}

The Common Public Radio Interface (CPRI) specification was issued by a consortium of radio equipment manufacturers with the aim of standardizing the communication interface between CU and RUs\footnote{The terminology used in CPRI is Radio Equipment Control (REC) and Radio Equipment (RE), respectively.} on the fronthaul network. CPRI prescribes, on the one hand, the use of sampling and scalar quantization for the digitization of the baseband signals, and, on the other, a constant bit rate serial interface for the transmission of the resulting bit rate. Note that the baseband signals are either obtained from downconversion for the uplink or produced by the CU after baseband processing (see next section) for the downlink.
The CPRI interface specifies a frame structure that is designed to carry user-plane data, namely the quantized IQ samples, along with the control and management plane, for, e.g., error detection and correction, and the synchronization plane data. It supports 3GPP GSM/EDGE, 3GPP UTRA and LTE, and allows for star, chain, tree, ring and multihop fronthaul topologies. CPRI signals are defined at different bit rates up to 9.8 Gbps and are constrained by strict requirements in terms of probability of error ($10^{-12}$), timing accuracy (0.002 ppm) and delay (5 ${\mu}s$ excluding propagation).

The line rates produced by CPRI are proportional to the bandwidth of the signal to be digitized, to the number of receive antennas and to the number of bits per sample, where the number of bits per I or Q sample is in the range 8-20 bits per sample for LTE in both the uplink and the downlink. Accordingly, the bit rate required for LTE base stations that serves multiple cell sectors with carrier aggregation and multiple antennas easily exceeds the maximum CPRI rate of 9.8 Gbs and hence the capacity of standard fiber optic links (see, e.g., \cite{IDT}). More discussion on CPRI can be found in Sec. \ref{sec:standard}.

\subsection{Point-to-Point Compression}\label{sec:ptp compression}

As discussed, the basic approach prescribed by CPRI, which is based on sampling and scalar quantization, is bound to produce bit rates that are difficult to accommodate within the available fronthaul capacities -- most notably for small cells with wireless fronthauling and for larger cells with optical fronthaul links in the presence of carrier aggregation and large-array MIMO transceivers. This has motivated the design of strategies that reduce the bit rate of the CPRI data stream while limiting the distortion incurred on the quantized signal. Here we provide an overview of these schemes by differentiating between techniques that adhere to the standard C-RAN implementation with full migration of baseband processing at the RU and solutions that explore different functional splits between RU and CU.

\subsubsection{Compressed CPRI}
In the first class, we have techniques that reduce the CPRI fronthaul rate by means of compression. The so called compressed CPRI techniques are based on a number of principles, which are briefly discussed in the following.\par
1) \textit{Filtering and downsampling} \cite{Samardzija,Guo}: As per the CPRI standard, the time-domain signal is oversampled. For instance, for a 10 MHz LTE signal a sampling frequency of 15.36 MHz is adopted. Therefore, a low-pass filter followed by downsampling can be applied to the signal without affecting the information content. \par
2) \textit{Per-block scaling} \cite{Samardzija,Guo}: In order to overcome the limitations due to the large peak-to-peak variations of the time-domain signal, per-block scaling can be performed. Accordingly, the signal is divided into subblocks of small size (e.g., 32 samples in \cite{Samardzija}) and rescaling the signal in each subblock is carried out so that the peak-to-peak variations in the block fit the dynamic range of the quantizer. \par
3) \textit{Optimized non-uniform quantization} \cite{Samardzija,Guo}: Rather than adopting uniform scalar quantization, the quantization levels can be optimized as a function of the statistics of the baseband signal by means of standard strategies such as the Lloyd-Max algorithm.\par
4) \textit{Noise shaping} \cite{Nieman}: Due to the correlation of successive baseband samples, predictive, or noise shaping, quantization techniques based on a feedback filter can be beneficial to reduce the rate of optimized quantization.\par
5) \textit{Lossless compression} \cite{Vosoughi}\footnote{Reference \cite{Vosoughi} in fact considers time-domain modulation and not OFDM but the principle is the same discussed here.}: Any residual correlation among successive quantized baseband samples, possibly after predictive quantization, can be further leveraged by entropy coding techniques that aim at reducing the rate down to the entropy of the digitized signal.\par
As a rule of thumb, compressed CPRI techniques are seen to reduce the fronthaul rate by a factors around 3 \cite{Dotsch}.

\begin{figure}[t]
\begin{center}
\vspace{0.7cm}
\epsfxsize=9cm \leavevmode\epsfbox{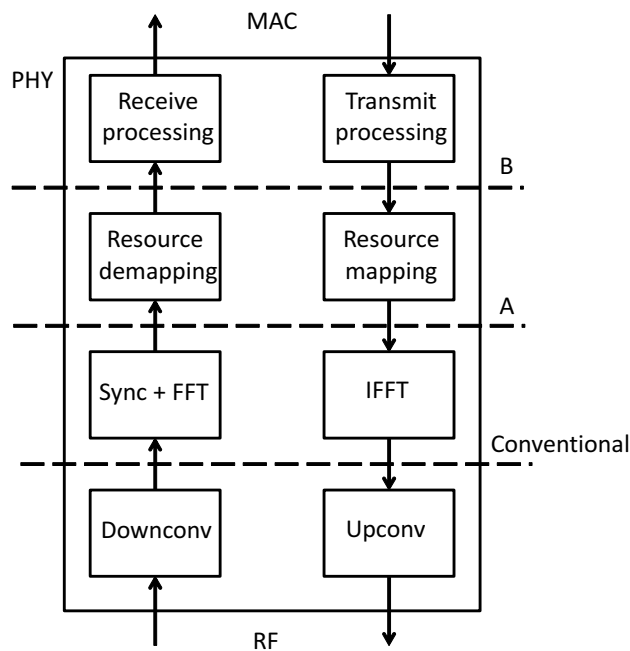} \caption{Alternative functional splits of the physical layer between CU and RU.} \label{fig:fsep}
\end{center}
\end{figure}

\subsubsection{Alternative Functional Splits}
In order to obtain further fronthaul rate reductions by means of point-to-point compression techniques, alternative functional splits to the conventional C-RAN implementation need to be explored \cite{Dotsch,Wubben}. To this end, some baseband functionalities at the physical layer (PHY), or Layer 1, can be implemented at the RU, rather than at the CU, such as frame synchronization, FFT/IFFT or resource demapping. Note that, while we focus here on the PHY layer, we discuss different functional splits at Layer 2 in Sec. \ref{sec:MAC}.

A first solution at the PHY layer prescribes the implementation of frame synchronization and FFT in the uplink and of the IFFT in the downlink at the RU (see demarcation "A" in Fig. \ref{fig:fsep}). The rest of the baseband functionalities, such as channel decoding/encoding, are instead performed at the CU. This functional split enables the signal to be quantized in the frequency domain, that is, after the FFT in the uplink and prior to the IFFT in the downlink. Given that the signal has a lower Peak-to-Average Ratio (PAPR) in the frequency domain, particularly in the LTE downlink, the number of bits per sample can be reduced at a minor cost in terms of signal-to-quantization-noise ratio. The experiments in \cite{Dotsch} do not demonstrate, however, very significant rate gains with this approach.

A more promising approach implements also resource demapping for the uplink and resource mapping for the downlink at the RU (see demarcation "B" in Fig. \ref{fig:fsep}). For the uplink, this implies that the RU can deconstruct the frame structure and distinguish among the different physical channels multiplexed in the resource blocks. As a result, the RU can apply different quantization strategies to distinct physical channels, e.g., by quantizing more finely channels carrying higher-order modulations. More importantly, in the case of lightly loaded frames, unused resource blocks can be neglected. This approach was shown in \cite{Dotsch,Lorca} to lead to compression ratios of the order of up to 30, hence an order of magnitude larger than with compressed CPRI, in the regime of small system loads. A similar approach is also implemented in the field trials reported in \cite{Grieger}.

\subsection{Network-Aware Compression}\label{sec:network-aware compression}
The solutions explored so far to address the problem of the excessive fronthaul capacity required by the C-RAN architecture have been based on point-to-point quantization and compression algorithms. Here we revisit the problem by taking a more fundamental viewpoint grounded in network information theory. As it will be discussed below, this network-aware perspective on the design of fronthaul transmission strategies has the potential to move significantly beyond the limitations of point-to-point approaches towards the network information-theoretic optimal performance.

\subsubsection{Uplink}
We start by analyzing the uplink. When taking a network-level perspective, a key observation is that the signals (\ref{eq:channel-uplink}) received by different RUs are correlated due to the fact that they represent noisy versions of the same signals $\mathbf{x}^{\mathrm{ul}}$. This correlation is expected to be particularly significant for dense networks -- an important use case for the C-RAN architecture. Importantly, the fact that the received signals are correlated can be leveraged by the RUs by implementing \textit{distributed source coding} algorithms, which have optimality properties in network information theory (see, e.g., \cite{ElGamal} for an introduction).

The key idea of distributed source coding can be easily explained with reference to the problem of compression or quantization with side information at the receiver's side. Specifically, given that the signals received by different RUs are correlated, once the CU has recovered the signal of one RU, that signal can be used as \textit{side information} for the decompression of the signal of another RU. This side information enables the second RU to reduce the required fronthaul rate with no penalty on the accuracy of the quantized signal. This process can be further iterated in a decision-feedback-type loop, whereby signals that have been already decompressed can be used as side information to alleviate the fronthaul requirements for the RUs whose signals have yet to be decompressed.

The coding strategy to be implemented at the RUs to leverage the side information at the receiver is known in information theory as \textit{Wyner-Ziv coding}. Note that Wyner-Ziv coding does not require the RU to be aware of the side information available at the CU but only of the correlation between the received signal and the side information.

Distributed source coding, or Wyner-Ziv coding, was demonstrated in a number of theoretical papers, including \cite{Sanderovich,dCoso,Park:TVT13,Zhou}, to offer significant potential performance gains. For example, in \cite{Park:CISS}, it was shown via numerical results to nearly double the edge-cell throughput for fixed average spectral efficiency and fronthaul capacities when implemented in a single macrocell overlaid with multiple smaller cells.

The implementation of Wyner-Ziv coding, including both quantization and compression, can leverage the mature state of the art on modern source coding (see, e.g., \cite{LDGM} and references therein). Nevertheless, an important issue that needs to be tackled is the need to inform each RU about the correlation between the received signal and the side information. This correlation depends on the channel state information of the involved RUs and can be provided by the CU to the RU. More practically, the CU could simply inform the RU about which particular quantizer/compressor to apply among the available algorithms in a codebook of possible choices. The design of such codebook and of rules for the selection of specific quantizers/compressors is an open research problem. A discussion on Wyner-Ziv coding using information theoretic arguments can be found in Sec. \ref{sec:baseband}.

\subsubsection{Downlink}
In the downlink, the traditional solution consisting of separate fronthaul quantizers/compressors is suboptimal, from a network information-theoretic viewpoint, based on a different principle, namely that of multivariate compression (see, e.g., \cite{ElGamal} for an introduction). \par
To introduce this principle, we first observe that the quantization noise added by fronthaul quantization can be regarded as a source of interference that affects the UEs' reception. With conventional point-to-point solutions, the CU has little control on this interfering signal given that the quantization/compression mapping is done separately for each RU. Multivariate, or joint, compression of the signals of all RUs overcomes this problem by enabling the shaping of the quantization regions for the vector of transmitted signals of all RUs. As a result, multivariate quantization/compression makes it possible to control the distribution of the quantization noise across multiple RUs in a similar way as precoding allows to shape the transmission of the useful signals across all the connected RUs. \par
The idea of multivariate compression for the C-RAN downlink was proposed in \cite{Park:TSP}. Moreover, recognizing that the quantization noise can be seen as an additional form of interference, reference \cite{Park:TSP} proposes to perform a joint design of precoding and multivariate compression using an information theoretic formulation. It was shown in \cite{Park:CISS} that multivariate compression yields performance gains that are comparable to distributed source coding for the uplink. \par
At a practical level, the implementation of multivariate compression hinges on the availability of channel state information at the CU, which is to be expected, and requires the CU to inform the RU about the quantization levels corresponding to each RU. As for the uplink, the resulting design issues are interesting open problems. Information theoretic considerations on multivariate compression can be found in Sec. \ref{sec:baseband}.

\section{\uppercase{Baseband Processing}}\label{sec:baseband}

As discussed in Sec. \ref{sec:introd}, one of the key advantages of the C-RAN architecture is that it provides a
platform for joint baseband signal processing across the multiple RUs in
both uplink and downlink. Such a cooperative network is often referred to as a
network MIMO or CoMP
\cite{Gesbert}. Joint transmission and reception
across the RUs allow the possibility for pre-compensation and subtraction of
interference across the cells. As inter-cell interference is the dominant
performance limiting factor in cellular networks, the C-RAN architecture can
achieve significantly higher data rates than conventional cellular networks.

As also seen, a key consideration in the design of cooperative coding strategies for C-RAN
is the capacity limit of the fronthaul. In this section, we elaborate on the
mathematical modeling of the compression process for a C-RAN system with
limited fronthaul and illustrate the effect of compression on baseband signal
processing by adopting an information theoretic framework.


The feasibility of cooperative joint signal processing in the C-RAN
architecture depends crucially on the ability of the RUs to obtain
instantaneous channel state information (CSI) and to precisely synchronize with
each other. In the uplink, timing differences can, in theory, be corrected for in
the digital domain, but downlink synchronization is imperative so that
signals transmitted by the different RUs are received synchronously at the
intended UE so as to achieve the cooperative beamforming effect. The rest of the section assumes the availability of CSI and the ability for the RUs to synchronize, and focuses on baseband beamforming design for inter-cell
interference mitigation. Furthermore, as described in Sec. \ref{sec:signal-model}, a flat-fading channel model is assumed
in order to illustrate the fundamental coding strategies in both uplink and
downlink.

\subsection{Uplink}

As introduced in Sec. \ref{sec:signal-model}, the C-RAN architecture consists of RUs that are partitioned into clusters, where each
cluster consists of $N_R$ RUs and is responsible for jointly decoding the
transmitted information from $N_U$ UEs. Let $M_R$ and $M_U$ be the number of
antennas in each of the RUs and the UEs respectively. The discrete-time baseband uplink C-RAN channel model can be written as (\ref{eq:channel-uplink}), which can be further detailed as
\begin{equation}
\left[
\begin{array}{c}
{\mathbf{y}}_{1}^{\mathrm{ul}} \\
{\mathbf{y}}_{2}^{\mathrm{ul}} \\
\vdots \\
{\mathbf{y}}_{N_R}^{\mathrm{ul}}
\end{array}
\right] =
\left[
\begin{array}{cccc}
{\mathbf{H}}_{1,1}^{\mathrm{ul}} & {\mathbf{H}}_{1,2}^{\mathrm{ul}} & \cdots &
	{\mathbf{H}}_{1,N_U}^{\mathrm{ul}} \\
{\mathbf{H}}_{2,1}^{\mathrm{ul}} & {\mathbf{H}}_{2,2}^{\mathrm{ul}} & \cdots &
	{\mathbf{H}}_{2,N_U}^{\mathrm{ul}} \\
\vdots & \vdots & \ddots & \vdots \\
{\mathbf{H}}_{N_R,1}^{\mathrm{ul}} & {\mathbf{H}}_{N_R,2}^{\mathrm{ul}} & \cdots &
	{\mathbf{H}}_{N_R,N_U}^{\mathrm{ul}}
\end{array}
\right]
\left[
\begin{array}{c}
{\mathbf{x}}_{1}^{\mathrm{ul}} \\
{\mathbf{x}}_{2}^{\mathrm{ul}} \\
\vdots \\
{\mathbf{x}}_{N_U}^{\mathrm{ul}}
\end{array}
\right]
+
\left[
\begin{array}{c}
{\mathbf{z}}_{1}^{\mathrm{ul}} \\
{\mathbf{z}}_{2}^{\mathrm{ul}} \\
\vdots \\
{\mathbf{z}}_{N_R}^{\mathrm{ul}}
\end{array}
\right],
\label{eq:ULchannel}
\end{equation}
where the noise terms ${\mathbf{z}}_{i}^{\mathrm{ul}}$ are assumed to
be additive white complex Gaussian vectors with variance $\sigma_{ul}^2$ on each
of its components. The goal of joint uplink processing is to utilize the received signals from all
the RUs, i.e., $\{ {\mathbf{y}}_{1}^{\mathrm{ul}},
{\mathbf{y}}_{2}^{\mathrm{ul}}, \cdots, {\mathbf{y}}_{N_R}^{\mathrm{ul}}\}$,
to jointly decode ${\mathbf{x}}_{1}^{\mathrm{ul}},
{\mathbf{x}}_{2}^{\mathrm{ul}}, \cdots, {\mathbf{x}}_{N_U}^{\mathrm{ul}}$.

The discussion in this section is restricted to a single-hop fronthaul topology, where
each RU $j$ is connected to the CU via a digital link of finite capacity.
If the fronthaul link capacity had been unlimited, the uplink channel
model would have been akin to a multiple-access channel with all the multiple
antennas across all the RUs being regarded as to form a single receiver.
In this case, well-known strategies such as linear receive beamforming and successive
interference cancellation (SIC) could be directly applied across the RUs to approach the
best achievable rates of such a multiple-access channel. In designing the
receive beamformers across the RUs, the minimum mean-square error (MMSE)
beamforming strategy, or a simpler zero-forcing beamforming strategy, can be
used, while treating multiuser interference as part of the background noise.

The coding strategy is considerably more complicated when the finite-capacity
constraints of the fronthaul links are taken into consideration. Toward this
end, as seen in Sec. \ref{sec:fronthaul compression}, the RUs must compress its observations and send a compressed version of
its IQ samples to the CU. From an information-theoretic viewpoint, the effect of compression can be modeled as
additional quantization noises (see, e.g., \cite{Gersho}). For example, if a
simple scalar uniform quantization scheme with $L$ quantization levels is used
for each I and Q component on each receive antenna, the quantization noise is approximately a uniform random
variable within the range $[-\frac{L}{2}, \frac{L}{2}]$. Assuming that the
maximum amplitude of the received signal in each of the antennas is within the interval
$[-\frac{M}{2}, \frac{M}{2}]$, the amount of fronthaul capacity needed to
support such uniform quantization is then $2\log_2(M/L)$ bits per sample, where the factor $2$ accounts for the I and Q components. Note
that the setting of the value $L$ provides a tradeoff between the fronthaul
capacity and the achievable rates. Intuitively, a coarser quantization, i.e.,
larger $L$, results in larger quantization noise, thus lower achievable rates,
but also less fronthaul. Conversely, finer quantization, i.e., smaller $L$,
results in higher achievable rates, but also requires more fronthaul capacity.

To capture such a tradeoff mathematically, and also to account for the fact that
vector quantization both across the antennas for each RU and across multiple
samples can be used, instead of scalar quantization, for higher
quantization efficiency, it is convenient to make the additional assumption
that the quantization noise can be modeled as an independent Gaussian process, i.e.,
\begin{equation}
{\mathbf \hat{y}}_j^{\mathrm ul} = {\mathbf y}_j^{\mathrm ul} + {\mathbf q}_j^{\mathrm ul}
\end{equation}
where ${\mathbf q}_j \sim {\mathcal CN}({\mathbf 0},{\mathbf Q}^{{\mathrm ul}}_j)$ and
${\mathbf Q}^{\mathrm{ul}}_j$ is an $M_R \times M_R$ covariance matrix
representing the compression of the received signals across $M_R$ antennas at
the $j$th RU. With this model of the compression process, the overall
achievable rate can now be readily written down as a function of the fronthaul
capacity.

To this end, assume that each of the UEs ${\mathbf x}_{i}^{\mathrm{ul}}$ transmits using a Gaussian
codebook ${\mathcal CN}({\mathbf 0}, {\mathbf \Sigma}_i^{\mathrm{ul}})$ with possibly
multiple data streams per user. In case of linear MMSE receive beamforming
across the RUs, the achievable rate for the $i$th UE can be expressed as:
\begin{eqnarray}
R_{i}^{\mathrm linear,ul} & \le & I({\mathbf x}_i^{\mathrm ul}; {\mathbf \hat{y}}_1^{\mathrm ul}
	\cdots {\mathbf \hat{y}}_{N_R}^{\mathrm ul}) \\
& = & \log \frac
{\left|\sum_{j=1}^{N_R} {\mathbf H}^{\mathrm{ul}}_{{\mathcal N}_U,j} {\mathbf \Sigma}_j^{\mathrm{ul}}
({\mathbf H}^{\mathrm{ul}}_{{\mathcal N}_U,j})^H + {\mathbf Q}_{{\mathcal N}_R}^{\mathrm{ul}} +
\sigma_{\mathrm ul}^2 {\mathbf I}\right|}
{\left|\sum_{j\neq i} {\mathbf H}^{\mathrm{ul}}_{{\mathcal N}_U,j} {\mathbf \Sigma}_j^{\mathrm{ul}}
({\mathbf H}^{\mathrm{ul}}_{{\mathcal N}_U,j})^H + {\mathbf Q}_{{\mathcal N}_R}^{\mathrm{ul}} +
\sigma_{\mathrm ul}^2 {\mathbf I}\right|}.
\label{eq:ULrate_linearBF}
\end{eqnarray}
When successive interference cancellation is used, assuming without loss of
generality a decoding order of the UEs as $1, 2, \cdots, N_U$, the achievable rate
for the $i$th user can instead be expressed as:
\begin{eqnarray}
R_i^{\mathrm SIC,ul} & \le & I({\mathbf x}_i^{\mathrm ul}; {\mathbf \hat{y}}_1^{\mathrm ul} \cdots
{\mathbf \hat{y}}_{N_R}^{\mathrm ul} |
{\mathbf x}_1^{\mathrm ul}, \cdots, {\mathbf x}_{i-1}^{\mathrm ul}) \\
& = & \log \frac
{\left|\sum_{j=i}^{N_R} {\mathbf H}^{\mathrm{ul}}_{{\mathcal N}_U,j} {\mathbf \Sigma}_j^{\mathrm{ul}}
({\mathbf H}^{\mathrm{ul}}_{{\mathcal N}_U,j})^H + {\mathbf Q}_{{\mathcal N}_R}^{\mathrm{ul}} +
\sigma_{\mathrm ul}^2 {\mathbf I}\right|}
{\left|\sum_{j=i+1}^{N_R} {\mathbf H}^{\mathrm{ul}}_{{\mathcal N}_U,j} {\mathbf \Sigma}_j^{\mathrm{ul}}
({\mathbf H}^{\mathrm{ul}}_{{\mathcal N}_U,j})^H + {\mathbf Q}_{{\mathcal N}_R}^{\mathrm{ul}} +
\sigma_{\mathrm ul}^2 {\mathbf I}\right|}.
\label{eq:ULrate_SIC}
\end{eqnarray}
In both cases, ${\mathbf H}^{\mathrm{ul}}_{{\mathcal N}_U,j}$
denotes the $j$th block-column of the matrix ${\mathbf H}^{\mathrm ul}$, i.e., the
collective channel from UE $j$ to all the RUs.
Note that the quantization process simply results in an additional noise term
in the rate expression, with the noise covariance matrix defined as
${\mathbf Q}_{{\mathcal N}_R}^{\mathrm{ul}} =
{\mathrm diag}({\mathbf{Q}}^{{\mathrm ul}}_1,\cdots,{\mathbf{Q}}^{{\mathrm ul}}_{N_R})$.

The above expression implicitly assumes that the quantization process is done
using point-to-point techniques, i.e., independently at each RU (see Sec. \ref{sec:fronthaul compression}). In this case, the amount of fronthaul capacity needed
to support such quantization at RU $j$ can be expressed based on rate-distortion
theory as (see, e.g., \cite{ElGamal})
\begin{eqnarray}
C_j^{\mathrm indep,ul} & \ge & I({\mathbf y}_j^{\mathrm ul} ; {\mathbf \hat{y}}_j^{\mathrm ul}) \\
& = & \log \frac{\left| \sum_{i=1}^{N_U} {\mathbf{H}}_{ji}^{\mathrm{ul}}
{\mathbf{\Sigma}}_{i}^{\mathrm{ul}} ({\mathbf{H}}_{ji}^{\mathrm{ul}})^H
+ \sigma_{\mathrm ul}^2 {\mathbf I} + {\mathbf{Q}}^{\mathrm{ul}}_j\right|}
{\left|{\mathbf{Q}}^{\mathrm{ul}}_j\right|}.
\label{eq:ULquantization_SU}
\end{eqnarray}
Alternatively, as discussed in Sec. \ref{sec:fronthaul compression}, Wyner-Ziv compression can be used to take advantage of the fact
that the compression of RUs can be done sequentially so that the compressed signals
of earlier RUs can act as the decoder side information for the compression of
later RUs. Assuming without loss of generality a decompression order of $1, 2, \cdots, N_R$
for the RUs, the fronthaul capacity constraint with Wyner-Ziv compression
can be shown to be
\begin{eqnarray}
C_j^{\mathrm WZ,ul} & \ge & I({\mathbf y}_j^{\mathrm ul} ; {\mathbf \hat{y}}_j^{\mathrm ul} |
{\mathbf \hat{y}}_1^{\mathrm ul}, \cdots, {\mathbf \hat{y}}_{j-1}^{\mathrm ul}) \\
& = & \log \frac{\left| {\mathbf{H}}_{{\mathcal J}_j {\mathcal N}_U}^{\mathrm{ul}}
{\mathbf{\Sigma}}_{{\mathcal N}_U}^{\mathrm{ul}} ({\mathbf{H}}_{
{\mathcal J}_j {\mathcal N}_U}^{\mathrm{ul}})^H
+ \sigma_{\mathrm ul}^2 {\mathbf I}_{{\mathcal J}_j} +
{\mathbf{Q}}^{\mathrm{ul}}_{{\mathcal J}_j}\right|}
{\left|{\mathbf{H}}_{{\mathcal J}_{j-1} {\mathcal N}_U}^{\mathrm{ul}}
{\mathbf{\Sigma}}_{{\mathcal N}_U}^{\mathrm{ul}} ({\mathbf{H}}_{
{\mathcal J}_{j-1} {\mathcal N}_U}^{\mathrm{ul}})^H
+ \sigma_{\mathrm ul}^2 {\mathbf I}_{{\mathcal J}_{j-1}} +
{\mathbf{Q}}^{\mathrm{ul}}_{{\mathcal J}_{j-1}}\right| \cdot
\left|{\mathbf{Q}}^{\mathrm{ul}}_j\right|},
\label{eq:ULquantization_WZ}
\end{eqnarray}
where we use the notations ${\mathcal J}_j = \{1,2,\cdots,j\}$ and
${\mathcal N}_U = \{1,2,\cdots,N_U \}$, and use
${\mathbf{H}}_{{\mathcal J}_{j} {\mathcal N}_U}^{\mathrm{ul}}$ to denote
the block-submatrix of ${\mathbf{H}}^{\mathrm{ul}}$ with indices taken from
${\mathcal J}_j$ and ${\mathcal N}_U$. Likewise,
${\mathbf{\Sigma}}_{{\mathcal N}_U}^{\mathrm{ul}}$ denotes a block-diagonal matrix
with block-diagonal entries
${\mathbf{\Sigma}}_{1}^{\mathrm{ul}}, \cdots, {\mathbf{\Sigma}}_{N_U}^{\mathrm{ul}}$; and a
similar definition applies to ${\mathbf{Q}}^{\mathrm{ul}}_{{\mathcal J}_{j-1}}$.

In summary, the uplink rate expressions (\ref{eq:ULrate_linearBF}) for linear
receive beamforming and (\ref{eq:ULrate_SIC}) for successive interference
cancellation provide information theoretical characterizations of the uplink
C-RAN capacity limit subject to fronthaul capacity constraints with either
independent per-link quantization (\ref{eq:ULquantization_SU}) or Wyner-Ziv
quantization (\ref{eq:ULquantization_WZ}).  These expressions implicitly
assumes the use of capacity and rate-distortion achieving codes, but the
performance with practical codes can also be easily obtained by incorporating
gap factors in the expressions (see e.g., \cite{Patel}). Also
implicit in the expressions is the decoding strategy at the CU of decoding the
compression codewords at the RUs first and then the transmitted codewords from the
UEs. Such a strategy has information theoretical justification
\cite{Sanderovich}, but we remark that this is not the only possible decoding strategy
(see e.g., \cite{Zhou13,Parkjoint}). Furthermore, as mentioned, the implementation of this strategy assumes
MMSE beamforming across the RUs.  The beamforming coefficients typically need to be
designed centrally at the CU as functions of the global CSI.

The achievable rate characterization points to the possibility that the
transmit covariance of the UEs and the quantization noise covariance at the RUs
may be jointly designed in order to maximize the overall system performance.
For example, a weighted rate-sum maximization problem may be formulated over
user scheduling, power control, transmit beamforming at the UEs, the
quantization noise covariance matrices at the RUs, and possibly the successive
compression and successive interference cancellation orders at the CU. Various
forms of this problem have appeared in the literature
\cite{dCoso,Zhou,ZhouGC,Park:TVT13}. The implementation of
the solutions to such an optimization, however, depends on the feasibility of
adaptive coding, modulation, and quantization codebooks, according to CSI,
scheduling, and user rates. As a first step for implementing C-RAN, fixed-rate
scalar uniform quantization is more likely to be used with quantization level set
according to the dynamic range of the analog-to-digital convertors and the
subsequent fronthaul capacity limits (see Sec. \ref{sec:fronthaul compression}). In fact, as shown in \cite{Zhou},
uniform quantization noise level is approximately optimal under suitable high
signal-to-noise ratio (SNR) conditions. In this case, the quantization noise simply
becomes additional background noise to be taken into consideration when
designing scheduling, power control, and receive beamforming strategies.

\subsection{Downlink}

In the downlink C-RAN architecture, baseband processing at the CU involves linear beamforming or non-linear techniques such as dirty paper coding that aim at ensuring that the signals transmitted by the RUs are received at the UE in such a
way that interference is minimized. If the fronthaul links between the RUs and
the CU have infinite capacities, the downlink C-RAN becomes a broadcast channel and standard network information theoretic results apply \cite{ElGamal}. The situation is instead more involved when the fronthaul links have finite capacities.
In this case, as seen, after the CU forms the
beamformed signals to be transmitted by the RUs, as functions of the user data
and CSI, such signals need to be
compressed before they can be sent to the RUs.

Mathematically, the discrete-time baseband downlink C-RAN channel model can be
written as (\ref{eq:channel-downlink}), or more specifically as
\begin{equation}
\left[
\begin{array}{c}
{\mathbf{y}}_{1}^{\mathrm{dl}} \\
{\mathbf{y}}_{2}^{\mathrm{dl}} \\
\vdots \\
{\mathbf{y}}_{N_U}^{\mathrm{dl}}
\end{array}
\right] =
\left[
\begin{array}{cccc}
{\mathbf{H}}_{1,1}^{\mathrm{dl}} & {\mathbf{H}}_{1,2}^{\mathrm{dl}} & \cdots &
	{\mathbf{H}}_{1,N_R}^{\mathrm{dl}} \\
{\mathbf{H}}_{1,2}^{\mathrm{dl}} & {\mathbf{H}}_{2,2}^{\mathrm{dl}} & \cdots &
	{\mathbf{H}}_{2,N_R}^{\mathrm{dl}} \\
\vdots & \vdots & \ddots & \vdots \\
{\mathbf{H}}_{N_U,1}^{\mathrm{dl}} & {\mathbf{H}}_{N_U,2}^{\mathrm{dl}} & \cdots &
	{\mathbf{H}}_{N_U,N_R}^{\mathrm{dl}}
\end{array}
\right]
\left[
\begin{array}{c}
{\mathbf{\hat{x}}}_{1}^{\mathrm{dl}} \\
{\mathbf{\hat{x}}}_{2}^{\mathrm{dl}} \\
\vdots \\
{\mathbf{\hat{x}}}_{N_R}^{\mathrm{dl}}
\end{array}
\right]
+
\left[
\begin{array}{c}
{\mathbf{z}}_{1}^{\mathrm{dl}} \\
{\mathbf{z}}_{2}^{\mathrm{dl}} \\
\vdots \\
{\mathbf{z}}_{N_U}^{\mathrm{dl}}
\end{array}
\right],
\label{eq:DLchannel}
\end{equation}
where ${\mathbf{z}}_{i}^{\mathrm{dl}}$ is the additive white complex Gaussian
vector with zero mean and variance $\sigma^2_{\mathrm dl}$ on each of its components.
Note that in a time-division duplex (TDD) system, the reciprocity of the uplink
and downlink channels would mean that
${\mathbf{H}}_{i,j}^{\mathrm{dl}} = {\mathbf{H}}_{j,i}^{\mathrm{ul}}$.

The transmit signals ${\mathbf{\hat{x}}}_{j}^{\mathrm{dl}}$ are quantized versions of the beamformed signals ${\mathbf{{x}}}_{j}^{\mathrm{dl}}$.
The quantization process can be modeled as the addition of quantization
noises as discussed above, yielding
\begin{equation}
{\mathbf{\hat{x}}}_{j}^{\mathrm{dl}} = {\mathbf{{x}}}_{j}^{\mathrm{dl}} +
{\mathbf q}_j^{\mathrm dl}.
\end{equation}
An interesting aspect of downlink quantization is that, in contrast
to uplink, where the quantization encoding in each RU is necessarily
independent, in the downlink the encoding operation is done centrally at the
CU, and thus \emph{correlated} quantization noises can be introduced. Such a
compression scheme is called multivariate compression, first introduced in the
C-RAN context in \cite{Park:TSP}, as discussed in Sec. \ref{sec:fronthaul compression}.

Let ${\mathbf s}_i^{\mathrm dl} \sim {\mathcal CN}({\mathbf 0}, {\mathbf \Sigma}_i^{\mathrm dl})$
be the beamformed signal intended for the $i$th UE to be transmitted across the
RUs, which may contain multiple data streams.  The eigenvectors of
${\mathbf \Sigma}_i^{\mathrm dl}$ are the transmit beamformers over the RUs.
As the desired transmit signal across the RUs is a combination of the intended
signals for all the $N_U$ UEs, i.e., ${\mathbf x}^{\mathrm dl} = \sum_{i=1}^{N_U} {\mathbf s}_i^{\mathrm dl}$, the transmit signal across the RUs is therefore
\begin{equation}
\left[
\begin{array}{c}
{\mathbf{{x}}}_{1}^{\mathrm{dl}} \\
{\mathbf{{x}}}_{2}^{\mathrm{dl}} \\
\vdots \\
{\mathbf{{x}}}_{N_R}^{\mathrm{dl}}
\end{array}
\right]
\sim
{\mathcal CN} \left({\mathbf 0}, \sum_{i=1}^{N_U} {\mathbf \Sigma}_i^{\mathrm dl} \right).
\end{equation}
In order to describe the quantization process, we rewrite components of the transmit
covariance matrix corresponding to each of the RUs separately as
\begin{equation}
\sum_{i=1}^{N_U} {\mathbf{\Sigma}}_i^{\mathrm dl}
=
\left[
\begin{array}{cccc}
{\mathbf{S}}_{1,1}^{\mathrm{dl}} & {\mathbf{S}}_{1,2}^{\mathrm{dl}} & \cdots &
	{\mathbf{S}}_{1,N_R}^{\mathrm{dl}} \\
{\mathbf{S}}_{1,2}^{\mathrm{dl}} & {\mathbf{S}}_{2,2}^{\mathrm{dl}} & \cdots &
	{\mathbf{S}}_{2,N_R}^{\mathrm{dl}} \\
\vdots & \vdots & \ddots & \vdots \\
{\mathbf{S}}_{N_R,1}^{\mathrm{dl}} & {\mathbf{S}}_{N_R,2}^{\mathrm{dl}} & \cdots &
	{\mathbf{S}}_{N_R,N_R}^{\mathrm{dl}}
\end{array}
\right],
\end{equation}
and also the quantization noise covariance as 
\begin{equation}
\left[
\begin{array}{c}
{\mathbf{{q}}}_{1}^{\mathrm{dl}} \\
{\mathbf{{q}}}_{2}^{\mathrm{dl}} \\
\vdots \\
{\mathbf{{q}}}_{N_R}^{\mathrm{dl}}
\end{array}
\right]
\sim
{\mathcal CN} \left({\mathbf 0},
\left[
\begin{array}{cccc}
{\mathbf{Q}}_{1,1}^{\mathrm{dl}} & {\mathbf{Q}}_{1,2}^{\mathrm{dl}} & \cdots &
	{\mathbf{Q}}_{1,N_R}^{\mathrm{dl}} \\
{\mathbf{Q}}_{1,2}^{\mathrm{dl}} & {\mathbf{Q}}_{2,2}^{\mathrm{dl}} & \cdots &
	{\mathbf{Q}}_{2,N_R}^{\mathrm{dl}} \\
\vdots & \vdots & \ddots & \vdots \\
{\mathbf{Q}}_{N_R,1}^{\mathrm{dl}} & {\mathbf{Q}}_{N_R,2}^{\mathrm{dl}} & \cdots &
	{\mathbf{Q}}_{N_R,N_R}^{\mathrm{dl}}
\end{array}
\right]
\right),
\end{equation}
where ${\mathbf{S}}_{i,j}^{\mathrm{dl}}$ and ${\mathbf{Q}}_{i,j}^{\mathrm{dl}}$
are $N_R \times N_R$ matrices.

If we use point-to-point fronthaul compression,  the quantization noises are independent and hence uncorrelated, i.e.,
${\mathbf Q}_{i,j}^{\mathrm dl} = {\mathbf 0}$ for $i \neq j$, and the fronthaul
capacity needed to generate ${\mathbf{\hat{x}}}_{j}^{\mathrm{dl}}$ is simply:
\begin{eqnarray}
C_j^{\mathrm indep,dl} & \ge & I({\mathbf x}_j^{\mathrm dl} ; {\mathbf \hat{x}}_j^{\mathrm dl}) \\
& = & \log \frac{\left| {{\mathbf{S}}^{\mathrm{dl}}_{jj}}
+ {\mathbf{Q}}^{\mathrm{dl}}_{jj}\right|}
{\left|{\mathbf{Q}}^{\mathrm{dl}}_{jj}\right|}.
\label{eq:DLquantization_SU}
\end{eqnarray}
If we instead utilize multivariate compression, as described in Sec. \ref{sec:fronthaul compression}, to generate correlated
quantization noises, extra fronthaul capacity would be needed. In a dual manner as in the uplink
Wyner-Ziv coding case, we restrict attention here to the performance achievable using successive encoding \cite{Park:TSP}. Without loss of generality, let the
encoding order of RUs be $1, 2, \cdots, N_R$. By simplifying the information theoretical expressions of \cite{Park:TSP}, it can be shown that the required fronthaul capacity can be expressed as follows:
\begin{eqnarray}
C_j^{\mathrm multi,dl} & \ge &
I({\mathbf x}_j^{\mathrm dl} ; {\mathbf \hat{x}}_j^{\mathrm dl}) +
I({\mathbf q}_j^{\mathrm dl} ; {\mathbf {q}}_1^{\mathrm dl}, \cdots,
	{\mathbf {q}}_{j-1}^{\mathrm dl}) \\
& = & \log \frac{\left| {{\mathbf{S}}^{\mathrm{dl}}_{jj}}
	+ {\mathbf{Q}}^{\mathrm{dl}}_{jj}\right|}
	{\left|{\mathbf{Q}}^{\mathrm{dl}}_{jj}\right|}
	+ \log \frac{\left| {{\mathbf{Q}}^{\mathrm{dl}}_{jj}}\right|}
	{\left|{\mathbf{Q}}^{\mathrm{dl}}_{jj} -
	{\mathbf{Q}}^{\mathrm{dl}}_{j{\mathcal J}_{j-1}}
	({\mathbf{Q}}^{\mathrm{dl}}_{{\mathcal J}_{j-1}{\mathcal J}_{j-1}})^{-1}
	{\mathbf{Q}}^{\mathrm{dl}}_{{\mathcal J}_{j-1}j} \right|},
\label{eq:DLquantization_Correlated}
\end{eqnarray}
where ${\mathcal J}_{j-1}=\{1, \cdots, j-1\}$ and
${\mathbf{Q}}^{\mathrm{dl}}_{{\mathcal J}_{j-1}{\mathcal J}_{j-1}}$
denotes the submatrix of the quantization covariance indexed by the subscripts,
and likewise for ${\mathbf{Q}}^{\mathrm{dl}}_{{\mathcal J}_{j-1}j}$.
Although generating correlated quantization noises requires extra fronthaul
capacity, as elaborated on in Sec. \ref{sec:fronthaul compression}, multivariate compression brings the advantage that the effective total noise at the UEs
may be lowered as the correlated quantization noises at the RUs can
potentially cancel each other after going through the channel, thus
improving the overall user rates for the system.

When multiuser interference is treated as noise, the achievable downlink rate
can be expressed as a function of the quantization noise covariance as
\begin{eqnarray}
R_i^{\mathrm linear,dl} & \le & I({\mathbf s}_i^{\mathrm dl}; {\mathbf y}_i^{\mathrm dl}) \\
& = & \log \frac
{\left|\sum_{j=1}^{N_U} {\mathbf H}^{\mathrm{dl}}_{i,{\mathcal N}_R}
\left( {\mathbf \Sigma}_j^{\mathrm{dl}} + {\mathbf Q}_{{\mathcal N}_R}^{\mathrm{dl}} \right)
({\mathbf H}^{\mathrm{dl}}_{i,{\mathcal N}_R})^H
+ \sigma_{\mathrm dl}^2 {\mathbf I}\right|}
{\left|\sum_{j\neq i} {\mathbf H}^{\mathrm{dl}}_{i,{\mathcal N}_R}
\left( {\mathbf \Sigma}_j^{\mathrm{dl}} + {\mathbf Q}_{{\mathcal N}_R}^{\mathrm{dl}} \right)
({\mathbf H}^{\mathrm{dl}}_{i,{\mathcal N}_R})^H + \sigma_{\mathrm dl}^2 {\mathbf I}\right|},
\label{eq:DLrate_linearBF}
\end{eqnarray}
where ${\mathcal N}_R = \{1, \cdots, N_R \}$, ${\mathbf H}^{\mathrm{dl}}_{j,{\mathcal N}_R}$
is the $j$th block-row of the channel matrix ${\mathbf H}^{\mathrm{dl}}$,
i.e., the collective channel from the RUs to UE $j$, and
${\mathbf Q}_{{\mathcal N}_R}^{\mathrm{dl}}$ is the quantization noise
covariance matrix across the $N_R$ RUs.
When dirty-paper coding is used in the downlink, multiuser interference can be
pre-subtracted. Assuming without loss of generality a successive precoding order of
UE $1, 2, \cdots, N_U$, the achievable rate for the $i$th user can be expressed as:
\begin{eqnarray}
R_i^{\mathrm DPC,dl} & \le & I({\mathbf s}_i^{\mathrm dl}; {\mathbf y}_i^{\mathrm dl}|
{\mathbf s}_1^{\mathrm dl}, \cdots, {\mathbf s}_{i-1}^{\mathrm dl}) \\
& = & \log \frac
{\left|\sum_{j=i}^{N_U} {\mathbf H}^{\mathrm{dl}}_{i,{\mathcal N}_R}
\left( {\mathbf \Sigma}_j^{\mathrm{dl}} + {\mathbf Q}_{{\mathcal N}_R}^{\mathrm{dl}} \right)
({\mathbf H}^{\mathrm{dl}}_{i,{\mathcal N}_R})^H +
\sigma_{\mathrm dl}^2 {\mathbf I}\right|}
{\left|\sum_{j=i+1}^{N_U} {\mathbf H}^{\mathrm{dl}}_{i,{\mathcal N}_R}
\left( {\mathbf \Sigma}_j^{\mathrm{dl}} + {\mathbf Q}_{{\mathcal N}_R}^{\mathrm{dl}} \right)
({\mathbf H}^{\mathrm{dl}}_{i,{\mathcal N}_R})^H + \sigma_{\mathrm dl}^2 {\mathbf I}\right|}.
\label{eq:DLrate_SIC}
\end{eqnarray}

In summary, the rate expressions (\ref{eq:DLrate_linearBF}) for linear
transmit beamforming and (\ref{eq:DLrate_SIC}) for dirty-paper coding provide
information theoretical characterization of the downlink C-RAN capacity limit
subject to fronthaul capacity constraints with either per-link quantization
(\ref{eq:DLquantization_SU}) or multivariate quantization
(\ref{eq:DLquantization_Correlated}). As above, the use of capacity and
rate-distortion achieving codes is assumed, but the expressions can be easily
modified to account for practical coding and compression methods.  These rate
characterizations again provide the possibility that the transmit covariance
intended for each UE and the quantization noise covariance at the RUs may be
jointly designed in order to maximize the overall system performance.
For example, a weighted sum-rate maximization problem may be formulated over
user scheduling, downlink power control, transmit beamformers, and the
quantization covariance setting at the RUs. Although this joint system-level
design problem is non-convex and fairly difficult to solve, algorithms capable
of achieving local optimum solutions have been devised for some forms of
this problem in \cite{Park:TSP,Patel}.

As in the uplink, the implementation of such solutions would require the use
of adaptive modulation, adaptive quantization, and the availability of global
CSI. Thus again, a first step for implementation of downlink C-RAN is likely
to involve simpler beamforming designs (such as zero-forcing) and scalar fixed
quantizers designed according to the per-antenna power constraints and the
fronthaul capacity limits.

\subsection{Alternative Functional Splits}

The discussion above assumes the standard C-RAN implementation in which the RUs are remote antenna heads tasked with
compression only and not with encoding and decoding of the UE data. As seen in Sec. \ref{sec:fronthaul compression}, this
functional split is preferred in C-RAN in order to make the RUs as simple as
possible, but it is not the only possible strategy. In terms of baseband processing, in the downlink, the CU may opt to share user data directly with the RUs,
instead of sharing the compressed version of the beamformed signals. The resulting replication of the UE data at multiple RUs yields an inefficient use of the fronthaul link capacity when the cooperation cluster size is large enough. However, data-sharing can be
effective when the fronthaul capacity is limited, i.e., when the cluster size
is relatively small (see, e.g., \cite{Kang}).  The optimization of the cooperation cluster is an
interesting problem, which has been dealt with extensively in the literature
\cite{Zhao13,Shi13,Zhuang14,Jun09,DaiYu_Access14,ZakhourGesbert11,MarschFettweis09}.

\section{\uppercase{Medium Access Control}}\label{sec:MAC}

In the previous sections, we have discussed cooperative techniques at the PHY layer that leverage the C-RAN architecture. As seen, these methods require the deployment of new infrastructure, including RUs and fronthaul link with tight capacity and latency constraints. It is, however, also of interest for mobile network operators to find solutions that reuse the existing infrastructure with the goal of cost-efficiently enhancing it with some centralized RAN functionalities. The implementation of an RU-CU functional split at Layer 2 is a promising candidate solution to achieve this goal, as it has been reported to drastically reduce the fronthaul requirements  -- up to factor $20$ depending on the system configuration \cite{SCF,NGMN} -- while still allowing for centralization gains by means of coordinated radio resource management (RRM). This section briefly reviews challenges and opportunities related to Layer 2 functional splitting.

\begin{figure}[t]
\begin{center}
\epsfxsize=8cm \leavevmode\epsfbox{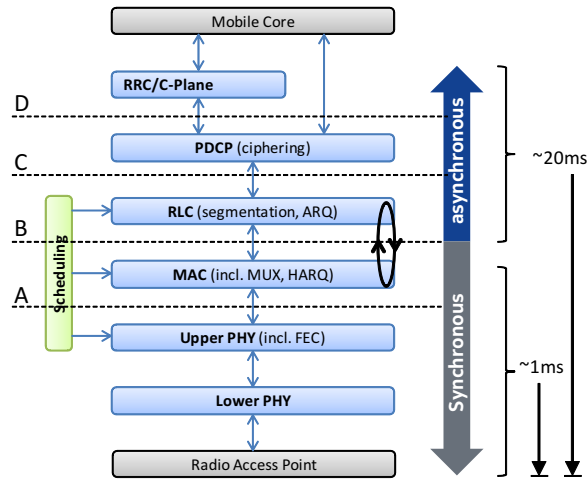} \vspace{0.7cm} \caption{Functional split options on RAN at Layer $2$.} \label{fig:stack}
\end{center}
\end{figure}

Fig.~\ref{fig:stack} shows several functional split options for Layer $2$ of the radio protocol stack. Note that, in the following, we use terminology based on the 3GPP LTE specifications. Since other technologies such as IEEE $802.16$ (WiMAX) have a similar radio architecture with functional equivalents, the discussion here applies in principle for them as well. The Layer $2$ is structured in sub-layers as follows \cite{TS36.300}:
\begin{itemize}
	\item Medium Access Control (MAC): this sub-layer is responsible for multiplexing and scheduling of control and user plane data into logical channels and transport blocks, and for hybrid automatic repeat request (HARQ) aimed at fast recovery from block errors;
	\item	Radio Link Control (RLC): this higher sub-layer is tasked with the segmentation of user data for the MAC scheduler, with buffering and with the ARQ protocol for improved link reliability;
	\item Packet Data Convergence Protocol (PDCP): this sub-layer, placed on top of the RLC sub-layer, is responsible for ciphering and integrity protection, for data forwarding aimed at hand-over, and for header compression on small data packets (e.g., voice data);
	\item Radio Resource Control (RRC): this sub-layer implements the control-plane protocol for radio link management and configuration, including measurements, admission control, and hand-over control.
\end{itemize}

An important aspect of Layer 2 sub-layers is the classification into \emph{synchronous} and \emph{asynchronous} protocols, which refers to the timing of the corresponding frame building process at the base station: synchronous protocols need to deliver data (e.g., transport blocks in case of MAC) within the LTE transmission time intervals (TTI) of $1$\,ms and hence have more stringent requirements on latency and jitter than the \emph{asynchronous} sub-layers, whose latency requirements are of the order of $20$\,ms.

The type and corresponding latency requirement, along with advantages and disadvantages, of different functional splits at Layer 2 are summarized in Table~\ref{tab:MACsummary}. In the rest of this section, we provide a more detailed discussion on these aspects.

\subsection{Constraints and Requirements}

The RU-CU split of Layer 2 functionalities is subject to specific constraints and requirements. To start, we observe that the fronthaul capacity requirements are less critical than for PHY-layer functional splits. In particular, the additional overhead of the control plane, e.g., signalling radio bearers, RRC messages, and protocol headers at Layer 2 adds approximately $10$\,\% to the overall bandwidth requirement as driven by user plane traffic, yielding for LTE to up to a theoretical maximum overall fronthaul rate of, e.g., $150$\,Mbps in downlink for a $20$\,MHz FDD system with $2$ transmit antennas \cite{SCF}. Note that this rate corresponds to worst-case traffic conditions, which should be considered when dimensioning the system, whereby the available radio resources are fully occupied and the highest modulation and coding scheme (MCS) (ordinal number $28$ in the 3GPP specifications) is used.

A first set of constraints arises from the fact that some functions are located in a single protocol layer, such as segmentation in RLC and ciphering in PDCP, while others span several protocol layers. This imposes implementation constraints on potential functional splits that involve the latter type of protocols, because information exchange and consequently signalling between RU and CU would be required if the corresponding functional split were implemented. An example is scheduling, which encompasses the upper PHY (for assigning transport blocks to resource blocks), the MAC sub-layer (for multiplexing and QoS scheduling) and the RLC sub-layer (for extracting the required number of bytes from corresponding buffers). In particular, the MAC scheduler needs to know the buffer occupancy at the RLC sub-layer in order to extract the selected number of bytes from the RLC radio bearer buffers according to the available radio resources and scheduled calculation. This process needs to be completed in a fraction of the TTI of $1$\,ms and involves a bi-directional information exchange, as indicated in Fig.~\ref{fig:stack}. As a consequence, barring a re-consideration of the protocol stack design of LTE, split B in Fig.~\ref{fig:stack} can be in practice ruled out as a potential candidate due to the discussed tight integration of MAC and RLC sub-layers.

\begin{figure}[t]
\begin{center}
\epsfxsize=8cm \leavevmode\epsfbox{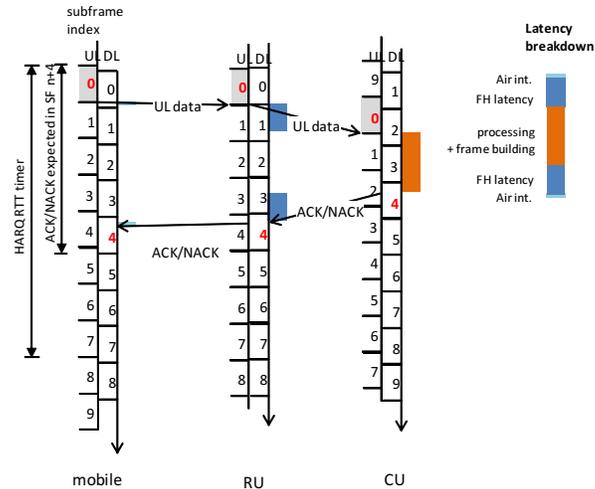} \vspace{0.7cm} \caption{HARQ timing in LTE systems.} \label{fig:harq}
\end{center}
\end{figure}

Other implementation constraints are determined by feedback loops involving the mobile device and by the related use of timers and procedures based on time-out events of some protocols. This is, for instance, the case for HARQ and ARQ, respectively at the MAC and RLC sub-layers, as well as for hand-over and connection control functions at the RRC sub-layer. Specifically, the HARQ feedback loop is the main constraining timer for all functional split options at Layer 2. As illustrated in Fig.~\ref{fig:harq}, the mobile device side expects an acknowledgement (negative or positive) in sub-frame $n+4$ counting from the sub-frame of the transmission in uplink \cite{TS36.213}. This imposes a limitation of below $3$\,ms on the round-trip time budget. As also indicated in the figure, this budget includes the round-trip transmission over the fronthaul as well as the processing and frame building at the CU. Assuming that sufficient processing power is available at the CU, this leads to a maximum tolerable fronthaul one-way latency of approximately ~$1$\,ms -- buffering for jitter not included. This requirement constitutes a challenge for functional split A. In the literature, only few papers have addressed this challenge so far, e.g., \cite{rost2014opportunistic,Dotsch,Khalili,Han1}. Above functional split A, timing requirements are less stringent. Specifically, ARQ and RRC timers are configurable, but in order to ensure the performance of key indicators such as hand-over failure and residual block error rate, a maximum latency in the range of $10$ to $20$\,ms for split options C and D should be assumed.


\subsection{Centralization Gains}

Having discussed the drawbacks related to requirements and implementation constraints of different functional splits at Layer 2, we now elaborate on their relative advantages in terms of centralization gains. Centralization gains can be classified into \emph{multiplexing gains}, which depend on the statistical properties of aggregated traffic and processing demand at the CU, and \emph{coordination gains}, which are due to coordinated radio resource management and control for a cluster of RUs. Note that the latter implies that the system implementation is capable of exchanging required information between protocol entities in the CU, which requires interfaces between functional entities (e.g., APIs).

Multiplexing gains for traffic aggregation apply mainly to the interfaces from the CU to the mobile core network. The reason is that the fronthaul needs to be capable of carrying the maximum possible throughput per RU for most splits. For splits C and D on asynchronous protocols, the fronthaul could be dimensioned also on a higher quartile of the bandwidth distribution (e.g., $99$\,\% quartile) with the risk of introducing some additional delay, or even additional packet losses \cite{iJOIND5.3}.  Coordination gains instead depend on the split option and the corresponding centralized functions, and two main cases can be distinguished:
\begin{itemize}
	\item \textit{Coordinated RRM}: this form of centralization includes scheduling, inter-cell interference coordination (ICIC), cell-based discontinuous transmission (DTX) and other techniques where the CU decides on the allocation and transmission of resources on a per-frame basis. The corresponding gains can be attained for split option A and below. For functional splits higher in the protocol stack, a centralized, more long term RRM approach is possible but would require additional signalling between the CU and the MAC entities in the RU (see, e.g.,~\cite{iJOIND3.3,FritzscheTWC,PateromichelakisAccess}).
	\item \textit{Centralized RRC}: this type of centralization amounts to the coordination of admission control, load balancing, hand-over parametrization and related self-organizing network (SON) functions which are executed on longer time ranges. These gains are possible for all split options, including split C and D.
\end{itemize}
A comprehensive overview of potential coordination gains depending on the functional split option is provided in \cite{iJOIND5.3}. Finally, there are some implicit gains such as centralization of ciphering in split option C, which implies that data is already protected for transport to the RUs and therefore does not need additional, and costly, transport layer security.

\subsection{Summary}

In summary, although various split options on Layer $2$ are possible in current cellular systems, the design of the protocol stack points strongly towards two main candidates, namely splitting below MAC (split A), which benefits from centralized RRM at the price of high requirements on backhaul latency, and a split between PDCP and RLC (split C), which is cost efficient and is already standardized as dual connectivity in LTE \cite{TS36.300}. As mentioned, Table~\ref{tab:MACsummary} provides an overview of the pros and cons of the four considered split options.
\begin{table}
	\centering
		\begin{tabularx}{\linewidth}{|l|l|l|X|X|X|}
			\hline
			Split	&	 Type	&	Fronthaul latency req.	&	Centralization gains & Pros	& Cons\\
			\hline
			\hline
			D	&	asynch.	&	$20$\,ms	&	Admission control, load balancing, SON functionality  &	Low req.\ on fronthaul and computational resources	& C-plane centralization only \\
			\hline
			C	&	asynch.	&	$20$\,ms	& D + moderate processing gains &	Low req., fronthaul ciphering included	&	Coordinated scheduling would require additional signaling\\
			\hline
			B	&	synch.	&	$\ll 1$\,ms	&	None &	None	&	Very high req. on latency, additional signalling required\\
			\hline
			A	&	synch.	&	$< 1$\,ms	&	C + centralized ICIC, scheduling &	High RRM gains possible	&	Higher req. on latency\\
			\hline			
		\end{tabularx}
	\caption{Summary of Layer 2 functional split options.}
	\label{tab:MACsummary}
\end{table}

\section{\uppercase{Radio Resource Management}}
\label{sec:RRA}

As discussed in the previous section, in a C-RAN with a functional split below the level A in Fig. \ref{fig:stack}, RRM may be carried out at the CU in a centralized fashion for the cluster of connected RUs. This centralized optimization is based on the available information at the CU, including queue state information, CSI and topological information about the fronthaul network. Due to the limitations of the fronthaul network and the need for possibly large-scale centralized optimization, RRM optimization in C-RANs offers significant technical challenges that are briefly reviewed, along with the state of the art on existing solutions, in this section. Specifically, we first discuss in Sec. \ref{sec:staticRRM} the static RRM problem that aims at maximizing performance metrics such as weighted sum rate in a given frame. Then, in Sec. \ref{sec:dynamicRRM}, we elaborate on the more general RRM problem of allocating resources across successive frames in a dynamic fashion by adapting to the available CSI and queue state information. As it will be discussed, solutions to the static problem often serve as components of the techniques addressing the dynamic scenario (see, e.g., \cite{neelybook}).

\subsection{Static RRM}\label{sec:staticRRM}

The static RRM problem amounts to the maximization of performance criteria such weighted sum-rate on a per-frame basis based on the available CSI. Below, we first discuss fully centralized solutions and then party decentralized approaches that leverage game-theoretic tools.

\subsubsection{Centralized Optimization}

The optimization of typical performance criteria, such as weighted sum-rate, amounts to non-convex, and possibly combinatorial, optimization problems with respect to the resource variables of interest, including downlink beamforming, uplink user association and RU clustering. Furthermore, these problems typically involves constraints that account for the limited fronthaul resources, such as the requirement to activate only a subset of RUs. We briefly review some approaches and solutions in the following.

Non-convexity with respect to the downlink beamforming variables is caused by the presence of inter-cell, or inter-RU, interference. This can be generally dealt with in various ways, most notably via successive convex approximation methods (see, e.g., \cite{Scutari}) and via techniques based on Fenchel-duality arguments or, equivalently, on the weighted minimum mean square error (WMMSE) method \cite{mi}. Instead, the mentioned fronthaul constraints are often formulated by introducing an $l_0$-norm regularization term in the objective function that enforces a penalty which is proportional to the number of active RUs, or, in other words, to the sparseness of the RU activation vector. To transform the corresponding non-convex problems into convex ones, standard $l_1$-norm approximation methods can be used to ensure sparsity of the resulting solution, or, more generally, mixed $l_1/l_p$-norm approximation techniques can be adopted to induce group sparsity (see, e.g., \cite{DaiYu_Access14}).

The approaches discussed above have been applied in the context of C-RAN in \cite{1}\cite{Shi13} with the aim of minimizing energy consumption; in \cite{3} for joint power and antenna selection optimization; in \cite{DaiYu_Access14} for weighted sum-rate maximization; and in \cite{5} for joint downlink precoding and uplink user-RU association optimization.

\subsubsection{Decentralized Optimization}

The centralized optimization discussed above requires the availability of CSI at the CU, which may impose a significant burden on the fronthaul, especially for large-scale C-RANs. To reduce this overhead, one may resort to decentralized solutions whereby the RUs self-organize into clusters based only on collected local information. To this end, the framework of coalition games can be adopted to develop cluster formation algorithm. This was proposed in reference \cite{VI-3-4}, which uses as utility function of a cluster the total data rate, and leverages a merge-split algorithm to obtain a stable cluster partition. A related work is \cite{VI-3-5}, in which interference from a legacy base station is considered that is coordinated with a coexisting C-RAN by means of a contract-based approach. Here, the proposed scheme aims at maximizing the utility of the C-RAN while preserving the performance of the legacy base station.

\subsection{Dynamic RRM}\label{sec:dynamicRRM}

While the solutions reviewed above operate on a per-frame basis, in practice, RRM needs to operate across multiple frames and to be adaptive to the time-varying conditions of the channel on the RAN and to the state of the queues. Dynamic RRM solutions that tackle this problem are reviewed in the following, by focusing on the two prominent approaches based on Markov decision processes (MDP) and Lyapunov optimization.

\subsubsection{Markov Decision Processes}

To elaborate, the system state of a C-RAN in a given frame can be generally characterized
by the current CSI and queue state information, which we denote as $\chi (t) =
[{\bf{H}}(t), {\bf{Q}}(t)]$, where $t$ represents the frame index, ${\bf{H}}(t)$ is the current CSI and the vector ${\bf{Q}}(t)$ describes the state of the queues. Under a Markovian model for the state $\chi (t)$, the dynamic RRM problem
can be modeled as a finite or infinite horizon average cost MDP. Under
proper technical conditions, this problem can be in principle solved by tackling the Bellman
equation. However, this approach incurs the curse of
dimensionality, since the number of system states grows exponentially with
the number of traffic queues maintained by the centralized CU. To
overcome this problem, the methods of approximate MDP, stochastic learning, and
continuous-time MDP could be used (see, e.g., \cite{Puterman}). The problem is even more pronounced in the absence of full state information, in which case the framework of Partially Observable MDPs (POMDPs), with its added complexity, needs to be considered.

A dynamic RRM solution that operates at both PHY and MAC layers has been proposed in\cite{JLiISJ} in the presence of imperfect CSI at the CU for the downlink by leveraging the POMDP framework. This reference proposes to reduce the complexity of the resulting solution by describing the trajectory of traffic queues by means of differential equations and hence in terms of a continuous-time MDP. In so doing, the value functions can be easily calculated using calculus, hence substantially reducing the computational burden.

\subsubsection{Lyapunov Optimization}

Lyapunov optimization provides another systematic approach for dynamic RRM optimization in C-RANs. Lyapunov optimization-based techniques are able to stabilize the queues hosted at the CUs while additionally optimizing some time-averaged performance metric \cite{neelybook}. The approach hinges on the minimization of the one-step conditional Lyapunov
drift-plus-penalty function
\begin{equation}
E[ L({\bf{Q}}(t + 1)) - L({\bf{Q}}(t))|{\bf{Q}}(t)] +
V{E}[g(t)| {{\bf{Q}}(t)}],
\end{equation}
where $L({\bf{Q}}(t)) = \frac{1}{2}\sum\nolimits_{i \in \mathcal
{I}} {{Q_i}{{(t)}^2}}$ is the Lyapunov function obtained by summing the squares of the queues' occupancies, $g(t)$ is the
system cost at slot $t$ and $V$ is a adjustable control parameter.

The dynamic RRM problem of network power consumption minimization by means of joint RU
activation and downlink beamforming was studied in \cite{PTeseng}\cite{JLiGC15} by leveraging the Lyapunov optimization framework. As shown therein, the resulting algorithm requires the solution of a static penalized weighted sum rate problem at each frame, which may be tackled as discussed above. Reference \cite{JLiGC14} includes also congestion control in the problem formulation and derives corresponding solutions based on Lyapunov optimization.

\section{\uppercase{System-Level Considerations}}

In this section, we provide a brief discussion on network architectures implementing C-RAN systems. We first discuss the basic architecture in Sec. \ref{sec:basic-architecture} and then briefly cover more advanced solutions in Sec. \ref{sec:advanced-architecture}.

\subsection{C-RAN Network Architecture}\label{sec:basic-architecture}

\begin{figure}[htb]
	\centering
	\epsfxsize=10cm \leavevmode\epsfbox{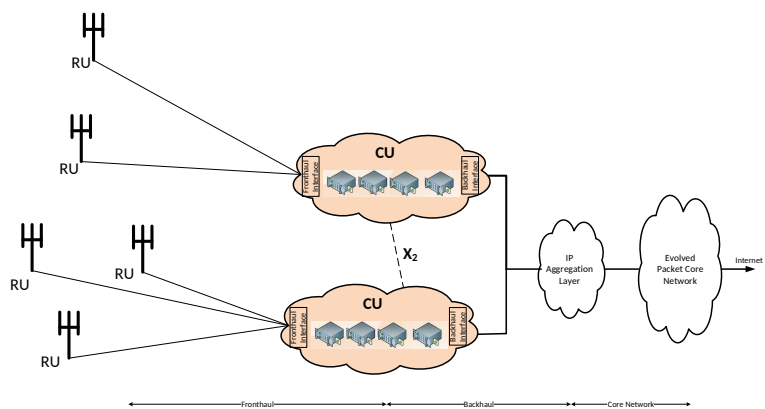}
	\caption{C-RAN system architecture.}
	\label{fig:onur}
\end{figure}

Fig. \ref{fig:onur} shows the basic architecture of a C-RAN system, which consists of the access, fronthaul, backhaul and packet core segments. In this architecture, the cell sites in the access network are connected to the cloud center, or CU, through fronthaul links. As discussed in Sec. \ref{sec:fronthaul compression} and Sec. \ref{sec:MAC}, the RUs may implement different functionalities at Layer 1 and Layer 2. In the most basic deployment, the RUs only perform RF operations, such as frequency up/down conversion, sampling and power amplification. In this case, the RUs contain the antennas and RF front-end hardware as well as the fronthaul interface software, e.g., CPRI, to communicate with the CU. Possible additional functionalities at Layer 1 and Layer 2 necessitate extra hardware and software modules at the RUs to coordinate and communicate with the CU (see, e.g., \cite{NGMN}).

The transport technology used in the fronthaul affects, and depends on, parameters such as cost, latency and distance between the radio sites and the Cloud Center. Fiber-optic and microwave links are the leading transport media for fronthauling, encompassing a large percentage of existing C-RAN developments \cite{NGMN}. Dedicated fiber solutions between the Cloud Center and RUs provide significant performance in terms of data rate and latency, but have been met with limited deployment due to cost associated with it. Optical Transport Networks (OTN), along with Wavelength Division Multiplexing Networks (WDM), provide high spectral efficiency by enabling fiber sharing among different cell sites with bidirectional transmission between the RUs and Cloud Center.

In the CU, multiple baseband units may be collocated that coordinate for the execution of the operations virtualized by the RUs in the access networks. A key design challenge for Cloud Centers is the development of cost-effective and high performance baseband pooling platforms, which may use, as further discussed in \cite{Checko}, either Digital Signal Processing (DSP) or General Purpose Processors (GPP) technologies. In addition to processing related to network access, the CU is also responsible for the interaction with the backbone network, e.g., Evolved Packet Core in LTE, via the backhaul (see Fig. \ref{fig:onur}). Current C-RAN technology solutions and trials mainly consider the fronthaul and backhaul segments separately, as in the recent platforms presented by Huawei and Ericsson \cite{NGMN}. In this case, fronthaul signals, which adhere to the serial CPRI transport format, are translated into Ethernet-based backhaul transport signals at the Cloud Center for transmission on the backhaul.

Due to baseband pooling and to the constraints on the I/Q data transmission in the fronthaul segment, the C-RAN architecture does not utilize the X2 interface to the extent of existing LTE systems. However, as shown in Fig. \ref{fig:onur}, C-RAN clustering methods across multiple CUs are also considered to leverage coordinated transmission for wider areas \cite{Checko}. In this regard, we observe that, even though inter-CU coordination cannot be as efficient as intra-RU coordination due to latency and capacity limitations in the fronthaul, multiple network management techniques such as time/frequency resource silencing techniques may be implemented across CUs.

\subsection{Next-Generation C-RAN Network Architecture}\label{sec:advanced-architecture}
	
\begin{figure}[htb]
	\centering
	\epsfxsize=10cm \leavevmode\epsfbox{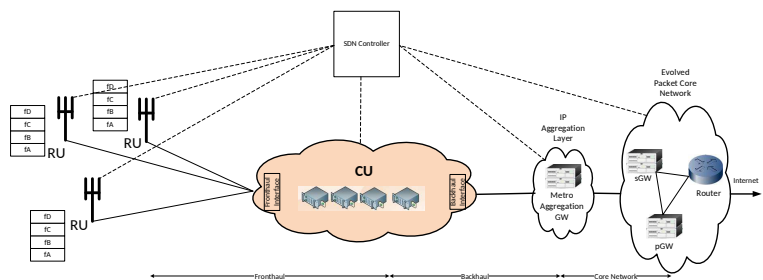}
	\caption{Flexible C-RAN architecture with fronthaul and backhaul segments.}
	\label{fig:stack1}
\end{figure}

In the state-of-the-art C-RAN architecture discussed above, it is generally quite complex, costly and inefficient to manage flexibly and dynamically the resources of the fronthaul, backhaul and core network segments. This is due to the heterogeneous technologies used for the corresponding network devices and their control elements. Recent advances in Software Defined Networking (SDN) technology, with its successful implementations such as OpenFlow, motivate the utilization of SDN network management tools for C-RAN deployments in order to overcome this limitation. Fig. \ref{fig:stack1} demonstrates a reference architecture that targets SDN-based unified network operation and transport mechanisms across the fronthaul, backhaul, and core network segments of a C-RAN architecture \cite{Liu,xhaul}.

In this architecture, a unified SDN-based control plane interfaces with the C-RAN network elements through dedicated control channels. Virtualized functions at the RUs, described in Sec. \ref{sec:fronthaul compression} (cf. Fig. 3) and identified in Fig. \ref{fig:stack1} for short as {fA,...,fD}, are dynamically coordinated by the SDN controller which assigns them to the corresponding nodes in the network. The function assignment procedure is based on network and link level abstractions at the network elements, which are populated through southbound and northbound interfaces and conveyed to the SDN controller. The abstracted parameters conveyed from the network elements to the controller through the northbound interface may include a wide range of inputs including link level conditions at the fronthaul such as bandwidth, signal to interference noise ratio, delay, etc. Similarly, the SDN controller, utilizing the southbound interface, conveys the configuration and execution information of the underlying virtual functions that are dynamically allocated per network element, e.g., RU, at a given network instance \cite{Liu}.

\section{\uppercase{Standardization}}\label{sec:standard}

The discussed potential performance gains and reduction in operating and maintenance cost offered by the C-RAN technology have resulted in significant industrial research and development efforts over the last decade. Similar to other incumbent telecommunication technologies and their life-cycles, the initial phase of C-RAN development was mostly led by the individual contributions and demonstrations of leading companies such as China Mobile \cite{China}, Huawei \cite{Huawei}, Ericsson \cite{NGMN}, Nokia Siemens Networks \cite{Guan}, and others. The subsequent, and ongoing, standardization efforts on C-RAN aim at developing a compatible fronthaul technology and its interfaces at the RUs and CU that enable multi-vendor operations. We briefly review below some of these activities.

As discussed in Sec. \ref{sec:fronthaul compression}, CPRI is currently the most widely deployed industry alliance standard that defines the specifications for the interface between the Radio Equipment Controller (REC) and the Radio Equipment (RE). CPRI defines a digitized I/Q transmission interface that supports serial, bidirectional and constant rate transmission on the fronthaul. The standard includes specifications for control plane, including strict synchronization and low-latency transmission via configured CPRI packetization, and for data plane \cite{CPRI}.

Open Base Station Architecture Initiative \cite{obsai} and Open Radio Interface \cite{ori} are other competing standards and industry associations that define interfaces and functional descriptions for the base station transceiver.

 Aiming at providing multi-tenancy support and dynamic functional allocation, hardware and functional virtualization have been discussed under the umbrella of NFV ISG in ETSI \cite{NFVISG}. Specifically, NFV ISG in ETSI targets a framework for telecom network virtualization that is directly applicable to the C-RAN architecture and aims at reducing the cost of deployment, at enabling multi-tenancy operation, and at allowing for easier operating and maintenance procedures.

\section{\uppercase{Concluding Remarks}}
This article has provided a short review of the state of the art and of ongoing activities in the industry and academia around the C-RAN technology. We have highlighted practical and theoretical aspects at Layer 1, including fronthaul compression and baseband processing; at Layer 2, with an emphasis on RU-CU functional splits; and at higher layers, including radio resource management. We have also discussed network architecture considerations and standardization efforts. Throughout the article, a tension has been emphasized between the two trends of virtualization, which prescribes wireless access nodes with only RF functionalities and entails significant capacity and latency requirements on the fronthaul architecture; and edge processing, which instead involves the implementation of a subset of Layer 1 and possibly also of Layer 2 functions at the edge nodes so as to reduce delays and alleviate architectural constraints. Ongoing activities point to solutions that find a balance between these two trends by means of flexible RAN and fronthaul/ backhaul technologies which allow the adaptation of the network operation to traffic type and system conditions.

\bibliographystyle{jcn}


\end{document}